\documentstyle[aps,amstex,amssymb,graphics,preprint]{revtex}
\begin{document}                                                       

\draft

\title {Fluctuation interaction of particles in a fluid}

\author{B.I.\ Ivlev}

\address{Department of Physics and Astronomy\\
University of South Carolina, Columbia, SC 29208\\
and\\
Universidad Aut\'onoma de San Luis Potos\'{\i}, 
Instituto de F\'{\i}sica, \\
San Luis Potos\'{\i}, S. L. P., 78000 Mexico}

\date{\today}
\maketitle

\begin{abstract}
The interaction of bodies in a fluid, mediated by hydrodynamic fluctuations and proposed by Dzyaloshinskii, Lifshitz, and 
Pitaevskii, is calculated exactly for parallel infinite planes and is shown to be attractive. The second mechanism of 
fluctuation interaction, proposed in the present work, is due to fluctuations of linear and angular velocities of bodies 
in a hydrodynamic medium and leads to a repulsion. The both mechanisms provide an interaction energy of two particles in a 
fluid of the order of temperature for the inter-particle distance of the micron scale, where the interaction mediated by 
electromagnetic fluctuations is small. When two particles approaches a wall, placed inside a fluid, the first (attractive)
interaction is violated a little, but the second (repulsive) weakens. This behavior correlates with the experimentally 
observed attraction of particles which appears when they approach a wall.

\end{abstract} \vskip 1.0cm


\narrowtext
\section{INTRODUCTION}
The interaction of colloidal particles is a matter of challenge and controversy despite of a long history of the
problem \cite{DER,ISR,RAJ}. An attraction of micron size particles separated by a distance of the micron scale has been 
established experimentally \cite{FRAD,GRIER,ARAUZ,JAIME}. An existence of the observable attraction at such big distance
sets a real puzzle. Colloidal particles accept in a fluid a surface charge screened by counterions at Debye's length 
and an important question is that can those particles with like charges attract each other due to solely electrostatic 
mean-field interaction. This type of attraction was predicted in Ref.~\cite{SOGAMI}. The results of Ref.~\cite{SOGAMI} are
not applicable to dielectric particles but only to those containing inside the electrolyte identical to outside one. The 
correct calculation of the interaction of that kind shows only repulsion in contrast to Ref. \cite{SOGAMI}. The 
electrostatic attraction between like-charge particles based on the mean-field approach was also predicted in Ref. 
\cite{BOWEN}, but that conclusion was incorrect, as shown in the work \cite{NEU}. An attraction of like charged particles
due to a mean-field mechanism seems to be extremely unlikely and the work \cite{NEU} provides strong arguments in support
of this statement. The conventional van der Waals attraction, mediated by the high-frequency (visible light) 
electromagnetic fluctuations, is very small at the micron distance consisting less than $10^{-2}T$ at room temperature
\cite{STAT} (see also Ref. \cite{AU}). Calculations of the effect of fluctuations of counterions 
\cite{MANN,BRUIN,LOWEN,TOKU} is actually an account of some sort of the van der Waals interaction mediated by low 
frequency (generic with the plasma frequency of the electrolyte) electromagnetic fluctuations. It is remarkably, the 
attraction found in the above works is sufficiently short range (compared with the micron length) and it is much bigger 
than temperature what enables to compete with the Coulomb repulsion on such short distance. This contrasts to the
experimental situation \cite{FRAD,GRIER,ARAUZ}, where the attraction appears on micron distance when the Coulomb 
repulsion becomes {\it of the order of temperature}. As one can conclude, neither an electrostatic mean-field interaction 
nor the van der Waals-type attraction, mediated by high-frequency \cite{STAT} and low-frequency 
\cite{MANN,BRUIN,LOWEN,TOKU} electromagnetic fluctuations, can explain the observed attraction 
\cite{FRAD,GRIER,ARAUZ,JAIME} of two colloidal particles separated by the micron scale distance.

The first goal of this work is to calculate exactly the interaction of two parallel infinite planes mediated by 
hydrodynamic fluctuations and to draw a conclusion about the analogous interaction of two particles. This type of 
fluctuation forces was discussed firstly by Dzyaloshinskii, Lifshitz, and Pitaevskii \cite{DZYAL}. It turns out, the 
interaction mediated by hydrodynamic fluctuations is attractive and not small at the micron distance and is relevant for 
explanation of the observed attraction between colloidal particles \cite{FRAD,GRIER,ARAUZ,JAIME}. The second goal of this 
work is to propose a new type of interaction of particles mediated by thermal fluctuations of linear and angular 
velocities (body variables) of particles in a hydrodynamic medium. This interaction is repulsive in contrast to one 
mediated solely by fluid fluctuations and in combination with it can explain the experimentally observed interaction of 
colloidal particles.
\section{HOW TO ACCOUNT THE HYDRODYNAMIC VAN DE WAALS INTERACTION} 
Suppose two bodies are placed inside a hydrodynamic medium, they are totally fixed in space, and serves only as obstacles 
for a fluid motion due to stick boundary conditions at their surfaces. There is no any macroscopic motion in the system 
and the only motion is caused by thermal fluctuations of the fluid velocity $\vec v(\vec r,t)$. In this case the free 
energy of thermal fluctuations of the fluid $F(R)$ depends on the distance $R$ between bodies. The function
\begin{equation}
\label{1}
U_{vdW}(R)=F(R)-F(\infty)
\end{equation}
is an interaction mediated by fluid fluctuations. Analogously to conventional van der Waals interaction mediated by
electromagnetic fluctuations, the potential (\ref{1}) can be called {\it the hydrodynamic van der Waals interaction}. 
To find the free energy of thermal fluctuations of the fluid one can start with the linearized Navier-Stokes equation 
\cite{HYDR}
\begin{equation}
\label{e1}
\rho\frac{\partial\vec v}{\partial t}=-\nabla p + \eta\nabla^{2}\vec v + \left(\zeta + \frac{\eta}{3}\right)\nabla\vec v
\end{equation}
The solution of Eq. \ref{e1} can be written in the form
\begin{equation}
\label{e2}
\vec v = {\rm curl}\hspace{0.1cm}\vec a + \nabla\phi
\end{equation}   
where the vector $\vec a$ satisfies the equation
\begin{equation}
\label{e3}
\rho\frac{\partial\vec a}{\partial t}=\eta\nabla^{2}\vec a
\end{equation}
The scalar $\phi$ is determined by the equation
\begin{equation}
\label{e4}
\rho\frac{\partial\nabla^{2}\phi}{\partial t}=-c^{2}\nabla^{2}\rho + \left(\zeta + \frac{4\eta}{3}\right)\nabla^{2}
(\nabla^{2}\phi)
\end{equation}
In Eq. \ref{e4} the schematic expression of $\nabla p$ through the sound velocity $c=\sqrt{\partial p/\partial\rho}$ is 
used. The density obeys the continuity equation $\partial\rho/\partial t=-\rho\nabla^{2}\phi$ what, together with 
Eq. \ref{e4}, provides the linear equation for $\nabla^{2}\phi$. Eq. \ref{e3} sets the typical frequency of fluctuations 
$\eta k^{2}/\rho$, where $k$ is a wave vector. At this frequency the term $c^{2}\nabla^{2}\rho\sim \nabla^{2}\phi$ in Eq. 
\ref{e4} dominates for sufficiently small wave vectors $k<1/R_0$, where
\begin{equation}
\label{e5}
R_0 = \frac{\eta}{\rho c}
\end{equation}
is some characteristic distance. The domination of that term requires the condition 
\begin{equation}
\label{e6}
\nabla^{2}\phi = 0
\end{equation}
From the physical point of view it means that on distances bigger than $R_0$ the fluid fluctuations correspond to an 
incompressible medium and fluctuations of the longitudinal velocity can be neglected. The vector $\vec a$ can be expanded 
by means of eigenfunctions $\vec {\psi}_{i}(\vec r)$ 
\begin{equation}
\label{e7}
\vec a = \sum_{i}\frac{\partial s_{i}}{\partial t}\hspace{0.1cm}\vec {\psi}_{i}(\vec r)
\end{equation}
of the boundary problem formulated in the following way
\begin{equation}
\label{e8}
-\nabla^{2}\vec a = \mu\vec a,\hspace{1cm}({\rm curl}\hspace{0.1cm}\vec a+\nabla\phi)\mid_{B}\hspace{0.1cm}=0
\end{equation}
where $B$ denotes the total boundary of all participating bodies and $\phi$ satisfies Eq. \ref{e6}. Eq. \ref{e3} now 
becomes of the form
\begin{equation}
\label{e9}
\rho\hspace{0.1cm}\frac{\partial^{2}s_{i}}{\partial t^2}+\eta\mu_{i}\hspace{0.1cm}\frac{\partial s_i}{\partial t}=0
\end{equation}
Here $s_i$ are some independent displacements of the fluid chosen according to the above diagonalization procedure. Now 
one should write down the free energy associated with the set (\ref{e9}) of independent harmonic oscillators, what is done 
in the next Section.
\section{FREE ENERGY OF THE DAMPED OSCILLATOR}
Let a harmonic oscillator obeys the classic dynamic equation
\begin{equation}             
\label{e10}
\frac{\partial^{2}s}{\partial t^2}+\gamma\frac{\partial s}{\partial t}+\Omega^{2}s=0
\end{equation}
Due to thermal and quantum oscillations the free energy of this oscillator is not zero. In the under-damped limit 
$\gamma=0$
the expression of the free energy $T\ln\left(2\sinh(\hbar\Omega/2T)\right)$ is well known. For non-zero $\gamma$ a 
calculation of the free energy constitutes a typical problem of the field called {\it dissipative quantum mechanics}. The 
initial ideas of this field, steamed to Feynman \cite{FEYN}, have been essentially developed by Caldeira and Leggett 
\cite{CALDLEG} and other researchers, see the book \cite{LEGGETT} and references therein. One of the way to calculate the 
free energy associated with the damped harmonic oscillator (\ref{e10}) is to consider a harmonic elastic string attached to
the free particle in the potential $m\Omega^{2}s^{2}/2$. In the classic limit this provides the equation of motion 
(\ref{e10}) with the finite $\gamma$ due to a radiation friction. Then the total harmonic system ``particle and string'' 
can be diagonalized to determined eigenfrequencies $\omega_i$ which differ from eigenfrequencies $\omega^{0}_{i}$ of the 
string without a particle. The free energy of the dissipative particle is determined as
\begin{equation}
\label{e11}
F=T\sum_{i}\ln\left(2\sinh\frac{\hbar\omega_i}{2T}\right) - T\sum_{i}\ln\left(2\sinh\frac{\hbar\omega^{0}_{i}}{2T}\right)
\end{equation}  
As a result, the free energy of the dissipative oscillator (\ref{e10}) in the classic limit of high temperature can be 
expressed in the form \cite{BLATTER}
\begin{equation}
\label{e12}
F=\int^{\infty}_{0}\frac{d\omega}{2\pi}\hspace{0.1cm}\frac{2T}{\omega}\ln\left(\frac{\hbar\omega}{T}\right)
{\rm Im}\frac{\Omega^{2}+\omega^{2}}{\Omega^{2}-\omega^{2}-i\gamma\omega}
\end{equation}
The result (\ref{e12}) can be clarified in the following way. The mean squared displacement of the oscillator 
(\ref{e10}), according to the fluctuation-dissipation theorem, has the form 
\begin{equation}
\label{e13}
<s^2>\hspace{0.1cm}=\int^{\infty}_{-\infty}\frac{d\omega}{2\pi}<s^2>_{\omega}
\end{equation}
where
\begin{equation}
\label{e14}
<s^2>_{\omega}\hspace{0.1cm}=\frac{i\hbar}{m}\hspace{0.1cm}\frac{1}{\omega^{2}-\Omega^{2}+i\gamma\omega}
\hspace{0.1cm}\coth\frac{\hbar\omega}{2T}
\end{equation}
The internal energy $U$ is a sum of kinetic and potential energies
\begin{equation}
\label{e15}
U=\int^{\infty}_{-\infty}\frac{d\omega}{2\pi}\left(\frac{m\omega^{2}}{2}+\frac{m\Omega^{2}}{2}\right)<s^2>_{\omega}
\end{equation}
Eq.~\ref{e15} follows also from the exact theory. The free energy $F=U-TS$, where the entropy is 
$S=-\partial F/\partial T$, can be found solving the differential equation what results in the expression (\ref{e12}).
In our over-damped case (\ref{e9}) one should put $\Omega=0$ and in the limit of high temperature $\hbar\gamma\ll T$ the 
free energy (\ref{e12}) is reduced to 
\begin{equation}
\label{e16}
F=\frac{T}{2}\ln\frac{\hbar\gamma}{T}
\end{equation}
When the viscous drag coefficient $\gamma$ tends to zero, the entropy of a free motion goes to infinity and the free energy 
decreases with the reduction of $\gamma$, according to Eq. \ref{e16}. As follows from Eqs. \ref{e10} and \ref{e9}, the
analogue of $\gamma$ in Eq. \ref{e16} is $\eta\mu_{i}/\rho$ and the free energy of fluid fluctuations is the sum of free 
energies of independent oscillators
\begin{equation}
\label{e17}
F=\frac{T}{2}\sum_{i}\ln\left(\frac{\hbar\eta}{T\rho}\mu_{i}\right)
\end{equation}
For fixed bodies the total number of fluid modes is not a fluctuating variable. The eigenvalues $\mu_i$ depend on a 
distance between bodies giving rise to an $R$-dependence of $F$, what determines the hydrodynamic van der Waals
interaction according to Eq.~\ref{1}. 
\section{EIGENVALUES FOR THE PLANE GEOMETRY}
Below we find the van der Waals interaction energy for two infinite plates immersed into a fluid, parallel to
$\{x,y\}$-plane and separated by the distance $R$. The $z$-axis is cut by planes at points $z=\pm R/2$. The solution of 
Eq.~\ref{e8} with the definition $k_{z}=\sqrt{\mu-k^2}$ has the form 
\begin{equation}
\label{e18}
\vec {a}(\vec {r},z)=\int\frac{d^{2}k}{(2\pi)^2}\left(\vec {A}_{k}\cos zk_{z}+\vec {B}_{k}\sin zk_{z}\right)
\exp{(i\vec k\vec r)}
\end{equation}
where $\vec r=\{x,y\}$. The solution of Eq. \ref{e6} can be presented as
\begin{equation}
\label{e19}
\phi(\vec r,z)=\int\frac{d^{2}k}{(2\pi)^2}\left(\alpha_{k}\cosh kz + \beta_{k}\sinh kz\right)\exp{(i\vec k\vec r)}
\end{equation}
The boundary condition (\ref{e8}) at $z=\pm R/2$ results in six equations for six independent variables 
$\alpha, \beta, A_{z}, B_{z}, (k_{y}A_{x}-k_{x}A_{y})$, and $(k_{y}B_{x}-k_{x}B_{y})$. There are two sets of eigenvalues 
determined by the following relations
\begin{equation}
\label{e20}
k_{z}\tanh\frac{kR}{2}-k\tan\frac{Rk_{z}}{2}=0\hspace{0.1cm},\hspace{1cm}k_{z}\tan\frac{Rk_{z}}{2}+k\tanh\frac{kR}{2}=0
\end{equation}
The solution of Eq. \ref{e20} can be written in the form
\begin{equation}
\label{e21}
k_{1z}=q_{1n}+\frac{2}{R}\alpha_{1n}\hspace{0.1cm},\hspace{1cm}k_{2z}=q_{2n}+\frac{2}{R}\alpha_{2n}
\end{equation}
where $q_{1n}=(1+2n)\pi/R$ and $q_{2n}=2n\pi/R$ with $n=1,2,3,...$
\section{FREE ENERGY OF THE FLUID BETWEEN TWO PLANES}
The eigenvalue $\mu=k^{2}+k^{2}_{z}$ should be inserted into Eq.~\ref{e17}. If the surface area of each plane is $S_0$,
the total free energy (\ref{e17}) can be expressed through the energy density per unit area of each plane 
$u_{vdW}=F/S_0$, which is given by the following relations
\begin{equation}
\label{e22}
u_{vdW}=W_{1}+W_{2}+W_{3}\hspace{0.1cm},\hspace{1cm}W_{1,2,3}=\frac{T}{2}\int\frac{d^{2}k}{(2\pi)^2}f_{1,2,3}
\end{equation} 
\begin{equation}
\label{e23}
f_{1}=\sum_{n\geq 1}\ln\frac{(k^{2}+k^{2}_{1z})(k^{2}+k^{2}_{2z})}{(k^{2}+q^{2}_{1n})(k^{2}+q^{2}_{2n})}
\end{equation}
\begin{equation}
\label{e24}
f_{2}=-\ln\left(\frac{\hbar\eta}{\rho T}\left(k^{2}+\frac{\pi^2}{R^2}\right)\right)\hspace{0.1cm},\hspace{1cm}
f_{3}=\sum_{n\geq 1}\ln\left(\frac{\hbar\eta}{\rho T}\left(k^{2}+\frac{\pi^{2}n^{2}}{R^2}\right)\right)
\end{equation}
In the expressions for $W_{1,2,3}$ one has to omit $R$-independent terms (surface energy) and $R$-linear terms (volume
energy), which are not relevant for the interaction problem. In Eq.~\ref{e23} the summation over $n$ can be extended up to 
infinity. In the expression for $f_3$ the set $\{2n,(1+2n)\}$ is denoted as $n$ and the diverging sum should be cut off
at $k\sim \pi n/R\sim 1/R_0$. It is easy to see that at $R\gg R_0$ Eqs.~\ref{e22} and \ref{e24} produce the $R$-dependent
part of $W_2$ 
\begin{equation}
\label{e25}
W_{2}=\frac{\pi T}{4R^{2}}\ln\frac{R}{R_0}
\end{equation}
The quantity $f_3$ in Eq. \ref{e24} can be written as a sum $f_{3}=f^{(1)}_{3}+f^{(2)}_{3}$, where
\begin{equation}
\label{e26}
f^{(1)}_{3}=\ln\prod_{n\geq 1}\left(1+\frac{k^{2}R^{2}}{\pi^{2}n^{2}}\right)=\ln\frac{\sinh kR}{kR}\hspace{0.1cm},
\hspace{1cm}f^{(2)}_{3}=\ln\prod_{n\geq 1}\frac{n^{2}R^{2}_{0}}{R^2}
\end{equation}
Comparing to Eq.~\ref{e24}, in Eq.~\ref{e26} the sum of constants is omitted since it leads to a $R$-linear term according
to the rule $\sum_{n}\rightarrow R\int dk_{z}/2\pi$. The $n$-product in Eq.~\ref{e26} is restricted by $n<\pi/R_0$ where
compression fluctuations become important. In reality, an effective cut off of the contribution of transverse fluctuations
in $f^{(2)}_{3}$ is smooth, what can be described by the relation
\begin{equation}
\label{e27}
f^{(2)}_{3}=\ln\prod_{n\geq 1}\frac{n^{2}R^{2}_{0}}{\psi(nR_{0}/R)R^2}
\end{equation}
$\psi(x)$ is an even analytic function, $\psi(0)=1$, and $\psi(x)\rightarrow x^2$ at $x\rightarrow\infty$. Differentiation
of Eq.~\ref{e27} produces the expression
\begin{equation}
\label{e28}
\frac{\partial f^{(2)}_{3}}{\partial R}=-\frac{1}{2}\sum^{\infty}_{-\infty}\frac{1}{\psi(nR_{0}/R)R^{2}}\hspace{0.1cm}
\frac{\partial}{\partial R}\left(R^{2}\psi\left(\frac{nR_{0}}{R}\right)\right)+\frac{1}{R}
\end{equation} 
The $n$-sum in Eq.~\ref{e28} at $R\gg R_0$ can be substituted by an integration with the exponential accuracy
$\partial f^{(2)}_{3}/\partial R=1/R - c/R_0$, where the numerical constant is
\begin{equation}
\label{e29}
c=\int^{\infty}_{-\infty}dx\left(1-\frac{x}{2}\hspace{0.1cm}\frac{\psi\hspace{0.1cm}'(x)}{\psi(x)}\right)
\end{equation}
Finally, in the limit $R\gg R_0$
\begin{equation}
\label{e30}
f^{(2)}_{3}=-c\frac{R}{R_0}+\ln\frac{R}{R_0}+c_1
\end{equation}
where $c_1$ is a numerical constant. Eq.~\ref{e30} is derived neglecting the small parameter $kR_0$. An account of finite 
$kR_0$ leads only to $R$-independent corrections of the result (\ref{e30}). Use of Eqs.~\ref{e26}, \ref{e30}, and 
\ref{e22} results in the following expression
\begin{equation}
\label{e31}
W_{3}=\frac{T}{2}\int\frac{d^{2}k}{(2\pi)^2}\ln\left(1-\exp{(-2kR)}\right)
\end{equation} 
After integration Eq. \ref{e31} reads
\begin{equation}
\label{e32}
W_{3}=-\frac{\zeta(3)}{16\pi}\hspace{0.1cm}\frac{T}{R^2}
\end{equation}
\section{THE HYDRODYNAMIC VAN DER WAALS INTERACTION}
As follows from Eqs.~\ref{e21} and \ref{e23},
\begin{equation}
\label{e33}
f_{1}=\sum^{\infty}_{n=1}\left( \ln\left(1+\frac{4}{R^2}\hspace{0.1cm}\frac{Rq_{1n}\alpha_{1n}+
\alpha^{2}_{2n}}{q^{2}_{1n}+k^2}\right)+\ln\left(1+\frac{4}{R^2}\hspace{0.1cm}\frac{Rq_{2n}\alpha_{2n}+
\alpha^{2}_{2n}}{q^{2}_{2n}+k^2}\right)\right)
\end{equation}
The main contribution to $W_1$ comes from $k\gg 1/R$ , what allows to use the perturbation theory for 
$\alpha_{1n}=\alpha(q_{1n})$ and $\alpha_{2n}=\alpha(q_{2n})$. In the limit $kR\gg 1$ the function $\alpha(q)$ satisfies
the equation
\begin{equation}
\label{e34}
\left(q+\frac{2\alpha}{R}\right)\tan \alpha=-k
\end{equation} 
The solution of Eq. \ref{e34} can be presented in the form $\alpha(q)=\alpha_{0}+\alpha_{1}+\alpha_{2}+...$ where
\begin{equation}
\label{e35}
\alpha_{0}=-\arctan\frac{k}{q}\hspace{0.1cm},\hspace{0.5cm}
\alpha_{1}(q)=\frac{2k\alpha_{0}(q)}{(q^{2}+k^{2})R}\hspace{0.1cm},\hspace{0.5cm}
\alpha_{2}(q)=\left(1+\frac{q}{k}\arctan\frac{k}{q}\right)\frac{\alpha^{2}_{1}(q)}{\alpha_{0}(q)}
\end{equation}
In the limit $kR\gg 1$ in Eq. \ref{e33} one can change summation by integration according to the rule 
$\sum_{n}\rightarrow R\int dq/\pi$, what brings only an exponentially small error
\begin{align}
\label{e36}
f_{1}=&-\ln\left(1+\frac{4\pi\alpha(\pi/R)+4\alpha^{2}(\pi/R)}{\pi^{2}+k^{2}R^{2}}\right)-
\frac{1}{2}\ln\left(1+\frac{4\alpha^{2}(0)}{R^{2}k^{2}}\right)\nonumber\\
&+R\int^{\infty}_{-\infty}\frac{dq}{2\pi}\ln\left(1+\frac{4}{R^2}\hspace{0.1cm}\frac{\alpha(q)qR+
\alpha^{2}(q)}{q^{2}+k^{2}}\right)
\end{align}
The first (after a constant) non-vanishing term in $f_1$ is proportional to $1/R^{2}k^{2}$. The integration in Eq. 
\ref{e36} yields 
\begin{equation}
\label{e37}
f_{1}=\frac{5\pi^2}{6R^{2}k^2}
\end{equation}
The $k$-integration in Eq.~\ref{e22} is cut off at $k\ll 1/R_0$ and results in the logarithmically big term
\begin{equation}
\label{e38}
W_{1}=\frac{5\pi T}{24R^2}\ln\frac{R}{R_0}
\end{equation}
The total hydrodynamic van der Waals interaction energy per unit area of each plane with the logarithmic accuracy at 
$R\gg R_0$ can be obtained from Eqs.~\ref{e22}, \ref{e25}, and \ref{e38}
\begin{equation}
\label{e39}
u_{vdW}=-\frac{\pi}{24}\hspace{0.1cm}\frac{T}{R^2}\ln\frac{R}{R_0}
\end{equation}
Eq.~\ref{e39} is an exact analytic result for the case of two parallel infinite planes. The interaction (\ref{e39}) is
mediated by by fluctuations with the wave length much shorter than a typical micron diameter of particles. In 
Eq.~\ref{e39} the inter-plane distance $R$ should exceed the length $R_0$, otherwise at smaller $R$ longitudinal 
fluctuations also contribute to formation of interaction. The hydrodynamic van der Waals interaction is attractive due to
suppression of dissipation by the restriction of motion of the fluctuating fluid which is provided by stick boundary 
conditions. This is equivalent to an effective reduction of viscous drag coefficients of fluid modes and hence,
according to Eq.~\ref{e16}, to a reduction of the interaction energy when the planes approaches each other.

It is instructive to compare the obtained hydrodynamic van der Waals energy (\ref{e39}) with the analogous value
$u^{em}_{vdW}$ due to high frequency (visible light) electromagnetic fluctuations, what is the conventional van der Waals
interaction \cite{STAT}. With weakly dispersive in the visible spectrum dielectric constants of polystyrene walls 2.40
and water between them 1.77 one can obtain
\begin{equation}  
\label{39}
u^{em}_{vdW}\simeq -1.1\times 10^{-4}\hspace{0.1cm}\frac{\hbar c}{R^{3}}
\end{equation}
The small numerical coefficient in Eq.~\ref{39} is an essential feature of the conventional (electromagnetic) van der 
Waals interaction. This is exactly that what makes this interaction small at the micron distance. In contrast, the 
hydrodynamic van der Waals interaction (\ref{e39}) is {\it not proportional to a small coefficient} and at the distance
$R\simeq 1\mu{\rm m}$
\begin{equation}
\label{40}
\frac{u_{vdW}}{u^{em}_{vdW}}\simeq 10^{3}
\end{equation}
The estimate $R_{0}\simeq 10^{-3}\mu{\rm m}$ for water is used in Eq.~\ref{40}. Despite previous formulas refer to the 
case of infinite planes, the numerical coefficient in the hydrodynamic van der Waals interaction $U_{vdW}$ of two spheres
also should not be expected to be small. As one can see in the case of conventional electromagnetic van der Waals 
potential, the same reasons are responsible for smallness of the numerical coefficients for both planar and double sphere 
geometries \cite{NIN}. For micron size particles separated by micron distance, when $a\sim R\sim 1\mu{\rm m}$, the 
interaction can be estimated as $U_{vdW}\simeq u_{vdW}R^{2}$ resulting in the relation
\begin{equation}
\label{41} 
U_{vdW}\simeq -{\rm const}\times T 
\end{equation}
where the constant is of the order of unity.   
\section{THE PARADOX}
In case of conventional van der Waals interaction mediated by electromagnetic fluctuations the mean values of electric
and magnetic fields are zero $<\vec E>=<\vec H>=0$. The stress tensor $\sigma_{ik}$ for electromagnetic field is 
quadratic with respect to fields and hence the mean value $<\sigma_{ik}>$ is not zero. This makes the origin of the 
force due to electromagnetic fluctuations physically understandable. The situation with hydrodynamic fluctuation forces is
different. The mean value of a solution of the linearized Navier-Stokes equation is zero $<\vec v>=0$ (the same for a 
fluctuation part of $p$). The hydrodynamic stress tensor
\begin{equation}
\label{e40}
\sigma_{ik}=\eta\left(\frac{\partial v_{i}}{\partial r_{k}}+\frac{\partial v_{k}}{\partial r_{i}}\right)-p\delta_{ik}
\end{equation}
is linear in fluctuation variables, its fluctuation part is zero, and a fluctuation force has to be zero in this 
approximation. A non-zero contribution to $<\sigma_{ik}>$ can result from the nonlinear terms in the Navier-Stokes
equation neglected at the above approximation. Account of this non-linearity has been done in Ref.~\cite{JONES} and a 
finite fluctuation force has been obtained. This result was shown to be incorrect in Ref.~\cite{MUD}, where the exact
mean value of the stress tensor (\ref{e40}) was found to be zero on the base of exact non-linearity of the Navier-Stokes
equations. A conclusion of Ref.~\cite{MUD} was that a fluctuation interaction mediated by hydrodynamic fluctuations was
impossible. This contrasts with the result (\ref{e39}). What is wrong?

To understand the situation let us consider the linear chain of small  particles, connected by elastic springs, shown in 
Fig.~1 and described by the dynamic equation
\begin{equation}
\label{e41}
\frac{\partial^{2}u_{n}}{\partial t^{2}}=\frac{s^2}{a^{2}_{0}}(u_{n+1}+u_{n-1}-2u_{n})
\end{equation} 
where $s$ is the sound velocity and $a_0$ is the period. Two spatially fixed big particles substitute small particles, as
shown in Fig.~1. The system is elastic and the force acting on a big particle, placed on the site $n$ is
\begin{equation}
\label{e42}
F_{n}=\frac{ms^2}{a^{2}_{0}}(u_{n+1}-u_{n-1})
\end{equation}   
Here $m$ is the mass of a a small particle. The energy of fluctuation motion of small particles $U_{vdW}$ is determined by
self frequencies in the system like in Eq.~\ref{e11}. Two different positions of big particles, ``natural'' in Fig.~1a and 
``compressed'' in Fig.~1b, have identical self frequencies since the system is harmonic and hence
$U^{(a)}_{vdW}=U^{(b)}_{vdW}$. In this situation there is no van der Waals force, what can be understood since the mean
value of the force (\ref{e42}), which is linear in displacement, is zero. On the other hand, for another ``natural'' 
position (Fig.~1c) self frequencies differ from those in Figs.~1a and 1b and hence 
$U^{(a)}_{vdW}\neq U^{(c)}_{vdW}$ and the van der Waals force is non-zero. The reason for this is that a transition from
the position (a) to the position (c) in Fig.~1 cannot occur within a harmonic approximation, one should destroy some 
harmonic springs and rearrange them again in a different way. The linear expression for force (\ref{e42}) is not valid
to describe a transition from (a) to (c) and a real force is non-linear,  what makes its average finite even for 
$<u_{n}>=0$.

An analogous situation takes place in hydrodynamics. According to its derivation, Eq.~\ref{e39} is valid only for 
discrete $R=Na_{0}$, where $a_0$ is inter-atomic distance and $N$ is an integer number. The full dependence of $u_{vdW}$ 
on $R$ has a structure on the atomic scale corresponding to remove of discrete atomic layers from the inter-plane region.
The hydrodynamic expression for the stress tensor (\ref{e40}) is not valid at such short scale like Eq.~\ref{e42} cannot 
describe breaking of harmonic bonds. In contrast to smooth van der Waals potential mediated by electromagnetic 
fluctuations, the interaction mediated by hydrodynamic ones has a structure as a function of distance on the 
atomic scale superimposed on the smooth function (\ref{e39}). To some extend, this is analogous to observation of the 
structured interaction potential \cite{CROCKER}, where the role of atoms was played by small particles. The resulting 
statement is that the hydrodynamic expression for the stress tensor (\ref{e40}) cannot be used for calculation of 
fluctuation forces since it becomes non-linear (and contributes to those forces) at short distances where the hydrodynamic
approach is not valid. Hydrodynamic fluctuation forces should be calculated on base of energy like it was done above. The 
conclusion of Ref.~\cite{MUD} on absence of hydrodynamic fluctuation forces based on use of the hydrodynamic stress tensor
is incorrect.
\section{INTERACTION MEDIATED BY BODY FLUCTUATIONS}
Suppose two identical spheres of the radius $a$ are separated by the center-to-center vector $\vec R$ and immersed into a 
fluid. There is no any macroscopic motion in the system except one caused by thermal fluctuations of body variables:
linear velocities of spheres $\vec u_{1,2}(t)$ and their angular velocities $\vec \Omega_{1,2}(t)$. The hydrodynamic 
medium cannot fluctuate itself and their motion is only induced by fluctuations of body variables. The resulting free
energy of fluctuations of body variables $I(R)$ depends on the inter-body distance due to stick boundary conditions 
imposed on an induced fluid velocity. The energy $I(R)$ is formed by fluctuations of some characteristic frequency,
specified below, for which fluctuations of the inter-body distance is relatively small. This allows to characterize the 
energy $I$ by the inter-body distance $R$. In reality, fluctuation effects of a fluid (leading to the hydrodynamic van der
Waals interaction $U_{vdW}$) and of body variables cannot be separated and they together result in the total fluctuation 
energy $E(R)$ which is not the sum $U_{vdW}+I$ in general case. Since the interaction is formed by by fluctuations of
some typical frequency, this description is valid only for a time scale bigger than the inverse of that frequency. For 
shorter time scale the reduction of the interaction between bodies to a pairwise potential is not correct.

There is a special case, when the total interaction energy of two particles, mediated both by fluid fluctuations
$\vec v(\vec r,t)$ and body fluctuations $\vec u_{1,2}(t)$, can be written as the sum
\begin{equation}
\label{e43}  
E(R)=U_{vdW}+I(R)
\end{equation}
Here $U_{vdW}$ is contributed only by fluid fluctuations (bodies are at rest) and only body fluctuations are responsible 
for formation of $I$ (fluid fluctuations are only induced). The separation of different mechanisms (\ref{e43}) is valid 
when the mass density $\rho_0$ of particles is much bigger compared to the fluid density $\rho$. This limiting case of 
heavy particles is considered below what enables to understand the main physics and to draw a conclusion about the general
case $\rho_{0}\sim\rho$. As follows from the linearized Navier-Stokes equation (\ref{e1}), a typical fluctuation frequency
of the transverse fluid velocity (with ${\rm div}\vec v=0$) is
$\omega_{v}=\omega_{0}a^{2}/R^2$, where
\begin{equation}
\label{e44}
\omega_{0}=\frac{\eta}{\rho a^{2}}
\end{equation}
An estimate of a typical frequency of fluctuations of body variables follows from their
equation of motion. The linear velocities of two spheres obey the equations
\begin{equation}
\label{e45}
\frac{4\pi}{3}\hspace{0.1cm}\rho_{0}a^{3}\hspace{0.1cm}\frac{\partial\vec u_{1,2}}{\partial t}=\vec F_{1,2}
\end{equation}
The forces exerted on spheres by the fluid are expressed by means of the friction tensors $\zeta_0$ and $\zeta_1$
\begin{equation}
\label{e46}
F^{i}_{1,2}=-\zeta^{ik}_{0}u^{k}_{1,2}-\zeta^{ik}_{1}u^{k}_{2,1}
\end{equation}
Eq.~\ref{e45} with the estimate $\zeta_{0}\sim\pi\eta a$ gives a typical frequency due to the motion of body variables
$\omega_{u}=\omega_{0}\rho/\rho_{0}$. Under the condition
\begin{equation}
\label{e47}
\frac{R^{2}}{a^{2}}\ll \frac{\rho_{0}}{\rho}
\end{equation}
when the inequality $\rho\ll \rho_{0}$ holds, the typical frequency of body motion $\omega_{u}$ is much smaller than one
for motion of the fluid $\omega_{v}$. In the classic (only thermal fluctuations) case fluctuations of body variables, 
giving rise to $I(R)$, are well separated in frequency from fluctuations of the fluid velocity resulting in $U_{vdW}(R)$. 
On the short time scale $\omega^{-1}_{v}$ the distance $R$ fluctuates slowly and can be considered as a constant. On the
long time scale $\omega^{-1}_{u}$ the potential $U_{vdW}(R)$ is already formed but, due to the condition
$\partial^{2}U_{vdW}/\partial R^{2}\ll \rho_{0}a^{3}\omega^{2}_{u}$, it violates weakly the dynamics of body fluctuations,
which remains almost free. This leads to a possibility of separation of the two mechanisms in the total fluctuation energy
(\ref{e43}). At frequencies $\omega\sim\omega_{u}$, according to \cite{HYDR}, one can omit the frequency dispersion of 
forces in Eq.~\ref{e46} and the friction tensors, as shown in Ref.~\cite{MAZUR}, have the forms at $a\ll R$ 
\begin{equation}
\label{e48}
\zeta^{ik}_{0}=6\pi\eta a\left(\left(1+\frac{9a^{2}}{16R^{2}}\right)\delta_{ik}+\frac{27a^{2}}{16R^{4}}R_{i}R_{k}\right)
\end{equation}
and
\begin{equation}
\label{e49}
\zeta^{ik}_{1}=-6\pi\eta a\hspace{0.1cm}\frac{3a}{4R}\left(\delta_{ik}+\frac{R_{i}R_{k}}{R^{2}}\right)
\end{equation}
For the motion described by Eqs.~\ref{e45}, \ref{e46}, \ref{e48}, and \ref{e49} the Reynolds number
$(\rho a/\eta)\sqrt{<u^{2}>}$, where $<u^{2}>$ is thermal mean-squared velocity, is small. The $\zeta_{1}$-term in 
Eq.~\ref{e46} results in a macroscopic force on the body (1) due to a macroscopic motion of the body (2). This is 
traditionally called the hydrodynamic interaction \cite{LADD}.

The interaction $I(R)$ is formed on the time scale $1/\omega_{u}$, where the typical frequency of fluctuations of body 
variables can be estimated as $\omega\sim 10^{-6}{\rm sec}$ for a micron size particle in water. On this time scale 
thermal fluctuations of the inter-body distance are much smaller than micron range separation distance. This justifies
the concept of the pairwise potential $I(R)$ on the time scale bigger than $1/\omega_{u}$. Below we consider the 
inter-sphere distance $R$ satisfying inequalities
\begin{equation}
\label{e50}
1\ll\frac{R^{2}}{a^{2}}\ll\frac{\rho_{0}}{\rho}
\end{equation}
which unify the condition (\ref{e47}) of validity of Eq.~\ref{e43} together with the validity condition of the forms
(\ref{e48}) and (\ref{e49}) for friction tensors. Under the condition $a\ll R$ one can ignore fluctuations of angular 
velocities. Let us introduce the two vectors $\vec S_{1}$ and $\vec S_{2}$ proportional to the total and the relative
displacement of bodies
\begin{equation}
\label{e51}
\frac{\partial\vec S_{1,2}}{\partial t}=\frac{1}{\sqrt{2}}(\vec u_{1}\pm\vec u_{2})
\end{equation}
As follows from Eqs.~\ref{e45}, \ref{e46}, and \ref{e49}, the equations of motion for those displacements have the form
(there is no summation on the index $i$)
\begin{equation}
\label{e52}
\frac{4\pi}{3}\hspace{0.1cm}\rho_{0}a^{3}\hspace{0.1cm}\frac{\partial^{2}S^{i}_{1,2}}{\partial t^{2}}+
6\pi\eta a\lambda^{i}_{1,2}\hspace{0.1cm}\frac{\partial S^{i}_{1,2}}{\partial t}=0
\end{equation} 
where
\begin{equation}
\label{e53}
\lambda^{x}_{1,2}=\lambda^{y}_{1,2}=1+\frac{9a^{2}}{16R^{2}}\mp\frac{3a}{4R}\hspace{0.1cm},\hspace{2cm}
\lambda^{z}_{1,2}=1+\frac{9a^{2}}{4R^{2}}\mp\frac{3a}{2R}
\end{equation}
The vector $\vec R$ is supposed to direct along the $z$-axis. According to Eq.~\ref{e52}, there are six independent 
fluctuating coordinates, which contribute to the free energy. Comparing Eq.~\ref{e52} with the equation of motion
(\ref{e10}), one can write down, analogously to Eq.~\ref{e16}, the free energy of body fluctuations in the form
\begin{equation}
\label{e54}
F=3T\ln\left(\frac{9\rho}{2\rho_{0}}\hspace{0.1cm}\frac{\hbar\omega_{0}}{T}\right)+I(R)
\end{equation}      
where the interaction energy is
\begin{equation}
\label{e55}
I(R)=\frac{T}{2}\hspace{0.1cm}\sum^{3}_{i=1}\ln(\lambda^{i}_{1}\lambda^{i}_{2})
\end{equation} 
Insert of Eq.~\ref{e53} into Eq.~\ref{e55} produces the interaction energy due to body fluctuations
\begin{equation}
\label{e56}
I(R)=\frac{27}{16}\hspace{0.1cm}T\hspace{0.1cm}\frac{a^{2}}{R^{2}}
\end{equation}
The condition of applicability of Eq.~\ref{e56} is given by the criterion (\ref{e50}). The interaction energy (\ref{e56})
grows up under reduction of the inter-body distance. To clarify the physical meaning of this result one can consider a 
motion of the body, when the second body is at rest for simplicity. The friction force exerted by the fluid on the 
moving body is enhanced by the resting body since it stops the on its surface. This increases the viscous drag 
coefficient of the moving body with the reduction of $R$, what results, according to Eq.~\ref{e16}, in the {\it repulsive}
force caused by body fluctuations ($\partial I/\partial R <0$).
\section{DISCUSSION}
The total fluctuation energy of interaction of two particles in a fluid can be written as a sum of two different 
contributions (\ref{e43}) under the condition (\ref{e47}) when $\rho\ll\rho_0$. The repulsive part $I(R)$ is found under 
the condition (\ref{e50}). The van der Waals interaction $U_{vdW}(R)$ in Eq.~\ref{e43} (attractive part) is not
calculated exactly for the case of two spherical bodies, in this work only arguments for its estimation (\ref{41}) are 
given. The first remarkable feature of found fluctuation interactions is that they exceed much the conventional van der 
Waals interaction mediated by electromagnetic fluctuations \cite{STAT}, which is approximately $-(10^{-3}-10^{-2})T$ at
the surface separation of the order of $1\mu{\rm m}$. The second feature of these interactions is their different
characters: $I(R)$ is repulsive and $U_{vdW}(R)$ is attractive. The third feature is essentially different behavior of
these parts when two bodies are not in a bulk fluid but approach a wall \cite{GRIER} or are confined between two walls
\cite{ARAUZ}. The interaction $I(R)$ drops down while approaching a wall due to a reduction of the velocity field, 
transmitted from one body to another, what is caused by the stopping action of the wall (we do not consider in details how 
$I(R)$ is modified close to a wall). In contrast to this, the van der Waals part $U_{vdW}(R)$ decreases hardly near 
a wall since it is formed mainly by fluctuations of short wave lengths, of the order of $R_{0}$, which are weakly affected
by approaching a wall. In the general case ($\rho_{0}\sim\rho$) the separation of contributions (\ref{e43}) does not hold 
but, like in the limit $\rho\ll\rho_{0}$, the contribution of body fluctuations provides a repulsive effect in the total 
interaction energy $E(R)$ and fluid fluctuations give the attractive van der Waals contribution. While approaching a wall 
the repulsive contribution reduces but the attractive is violated a little. This seems to be relevant for explanation of 
the observed appearance of attraction between two bodies when they approach a wall \cite{GRIER}. In experiments 
\cite{FRAD}, \cite{GRIER}, and \cite{ARAUZ} the attraction potential was of the order of temperature what also correlates 
with the developed theory. For a detailed comparison with experiments further calculations based on the hydrodynamic 
approximation are required. In the work \cite{GRIER}, when two bodies approach a wall very close, the attraction 
disappears again. It can be explained by an increase of the Coulomb repulsion due to the electrostatic interaction through
the dielectric wall instead of Debye's screening.

The results of different experimental groups reveal a very stable feature of the particle interaction in a fluid: a value 
of the attractive potential is always of the order of temperature, what is a strong indication of a fluctuation nature
of the attraction. In the works \cite{FRAD} and \cite{ARAUZ} particles were confined between close glass walls, in the 
work \cite{GRIER} only one glass wall was used, and in the works \cite{JAIME} and \cite{JAIM} particles were placed on 
the air-water interface. This variety of situations enables to reject some mechanism of attraction. For example,  
mechanisms related to a charge of the glass walls unlikely play a crucial role, since the air-water interface in the 
experiments \cite{JAIME} and \cite{JAIM} essentially differs from the glass-water surface, nevertheless the value of the 
attraction potential was approximately the same like in the works \cite{FRAD}, \cite{ARAUZ}, and \cite{GRIER}. For 
particles on the air-water interface the concept of two types of interaction $I$ and $U_{vdW}$ also works, but it needs 
some modification.

In summary, the new mechanism of interaction of particles in a fluid is proposed based on positional and rotational
fluctuations of particles. This mechanism competes with the particle interaction mediated by hydrodynamic fluctuations
of a surrounding fluid. These interactions are not small at a micron size inter-particle distance and their interplay
correlates with the experimentally observed attraction between particles, which appears when they approach a wall inside 
a fluid.

I would like to thank C.-K. Au, A. Ladd, M. Medina-Noyola, and R. Rajagopalan for valuable discussions.


\begin{figure}[!ht]
\includegraphics{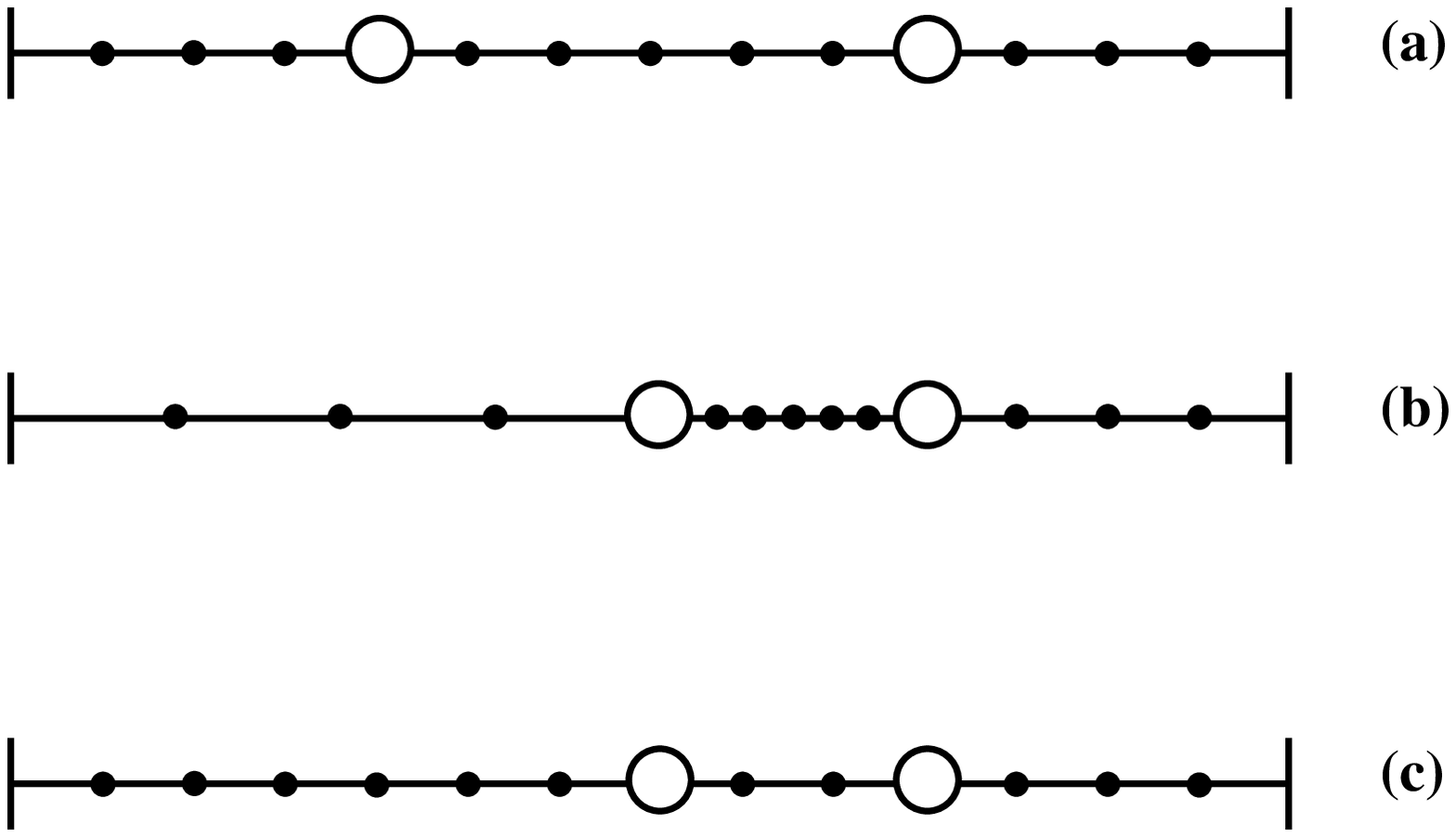}\\[8mm]
\caption{The linear chain of particles. Each horizontal segment between particles behaves like an elastic harmonic 
spring. (a) represents the ``natural'' positions of two attached particles (open circles), (b) is obtained from (a) 
only by a compression motion of attached particles without a destruction of harmonic bonds, (c) is another ``natural'' 
position obtained from (a) by destruction of harmonic bonds.}
\end{figure} 

%
%

\end{document}